# Tidal MerzA: Combining affective modelling and autonomous code generation through Reinforcement Learning

**Elizabeth Wilson[1] George Fazekas[1] Geraint Wiggins[2]**

[1]Queen Mary University of London, [2]Vrije Universiteit Brussel

**ABSTRACT**

This paper presents Tidal-MerzA, a novel system designed for collaborative performances between humans and a machine agent in the context of live coding, specifically focusing on the generation of musical patterns. Tidal-MerzA fuses two foundational models: ALCAA (Affective Live Coding Autonomous Agent) and Tidal Fuzz, a computational framework. By integrating affective modelling with computational generation, this system leverages reinforcement learning techniques to dynamically adapt music composition parameters within the TidalCycles framework, ensuring both affective qualities to the patterns and syntactical correctness. The development of Tidal-MerzA introduces two distinct agents: one focusing on the generation of mini-notation strings for musical expression, and another on the alignment of music with targeted affective states through reinforcement learning. This approach enhances the adaptability and creative potential of live coding practices and allows exploration of human-machine creative interactions. Tidal-MerzA advances the field of computational music generation, presenting a novel methodology for incorporating artificial intelligence into artistic practices.

## Introduction

This paper outlines the development of Tidal-MerzA, a system for collaborative performance with a machine agent in live coding, which merges insights from an affective model *ALCAA* (affective live coding autonomous agent) (Wilson et al. 2024) and computational generation framework *Tidal Fuzz* (Wilson et al. 2021). Its name is a portmanteau of “MERged Tidal-FuzZ and Alcaa” but also references the synonym for dada-ist practise invented by Kurt Schwitters (Shaffer 1990), to describe his collage and assemblage works.

Live coding is a term used to refer to performers creating art by writing computer code, usually in front of an audience (Collins 2003). In live coding, computer language is the primary medium for notation and describing the rules with which to synthesise artworks, in this case we consider the case where the output is musical pattern. The practice of live coding places a strong focus on the elements of liveness, embracing error, the use of random processes and clear mappings between syntax and output.

The TidalCycles live coding language is used to create autonomous patterns by Tidal-MerzA. TidalCycles is an expressive language, known for its flexibility and versatility in creating complex structural ideas, through its functional programming style and “mini-notation” syntax (McLean and Wiggins 2010). In Tidal-MerzA, these “mini-notation” strings—symbolic groupings to denote wider functions in TidalCycles—are a crucial aspect of representing and generating new patterns .

There have been different approaches to the task of autonomous code generation or agential design in live coding (Xambó 2021) and further to this, some attempts at analysing interactions with agents (Diapoulis 2023a) (Diapoulis 2023b). However, the role of affect is rarely incorporated into such models, where affective methodologies can add benefit as they investigate the affective processes that emerge (Magnusson 2023).

For this work, the aim is not only to generate TidalCycles code that is syntactically correct and evaluates to produce musical output, but also to consider the role of affective modelling of musical structural parameters and how to incorporate this into generation algorithms in a live composition setting. Affective modelling in machine learning provides a framework for collaborations with machines that create in a more human-like manner. However, the affective modelling process should acknowledge that machine aesthetics are often based on arbitrary metrics: where computers in lieu of the embodied emotional experience, would only discriminate or favour certain outcomes based on randomness and arbitrarily remove creative ideas from a conceptual space (Wiggins 2021), or otherwise if we accept computational aesthetics in any system we develop, then the method by which we generate them must be tied to a consciousness in the machine that we have yet to prove (Wilson et al. 2023). Instead, in Tidal-MerzA, the generation is guided by modelling affective response in humans.

The formation of a hybrid model that aims to combine affective capabilities with flexible coding is presented, offering a novel method for music generation blending the outcomes of previous work. The previous work on ALCAA presents a model of affect, translating literature findings into mathematical equations for creating music with specific affective qualities. However, a limitation of this model was its reliance on fixed code structures, restricting structural changes and underutilising TidalCycles' functional capabilities. Similarly, the previous work on the creation of the Tidal-Fuzz plugin introduced a model enabling the generation of syntactically correct code using various TidalCycles functions, by the process of a random walk through type signatures to produce executable code, but did not incorporate any modelling of affective equations. Tidal-MerzA aims to combine these two assets into one functional system.

For this hybrid system, two agents are consecutively built to attempt to capture all the dimensions outlined in the affective model-ALCAA. First, reinforcement learning (RL) techniques are used to dynamically adapt the parameters generated by the affective model within the flexible framework of TidalCycles. This framework enables the model to harness TidalCycles' extensive library of functions and patterns to generate music compositions that not only encapsulate desired emotional attributes but also adhere to the syntactical correctness of TidalCycles code. Secondly, specific mini-notation strings are produced that harness TidalCycles internal parsing of short-hand events.

In MerzA, the RL agent's actions correspond to the selection of musical elements within the TidalCycles framework. These actions are guided by a reward mechanism that evaluates the alignment between the generated music's affective attributes and the target affective states defined by the ALCAA model. Through trial and error, the agent refines its decision-making strategies, gradually learning which musical elements and TidalCycles functions to employ in order to evoke specific emotional responses. RL is particularly well-suited for this task because it allows the agent to learn and adapt in a dynamic environment and optimise its actions. In particular, modelling equations for different musical structural parameters were defined, namely: rhythmic structure, sound level/perceptual loudness, and tempo, modality, pitch register and pitch contour. Through the creation of two agents, these parameters are incorporated, preserving the original equations in this new mode of generation.

## The Reinforcement Learning Problem

As the integration of affective models forms the basis of this exploration, how these human affective states are modelled follows the valence-arousal model introduced by Russell (1980). Formerly, music psychology literature labelled affective states using a categorical model, suggesting these stem from a finite number of monopolar universal basic affects. However, currently various two or three- dimensional models have been more universally adopted, with Russell’s circumplex model of affect being commonly used, due to its ability to represent the complexities of affect. This approach employs valence (pleasure vs. displeasure) and arousal (high vs. low energy) as its dimensions, and is used in the research.

The reinforcement learning problem in the context of generating musical code based on valence-arousal coordinates involves training an agent to select sequences of code that correspond to desired affective qualities. The agent's goal is to maximise the cumulative reward received based on the affective quality of the generated musical code and its alignment with the specified valence-arousal coordinates. The agent interacts with an environment that provides feedback in the form of rewards, indicating how well the generated code matches the desired affective characteristics. By learning from this feedback, the reinforcement learning agent aims to discover a policy that maps valence-arousal coordinates to code sequences, enabling the generation of music that effectively captures the desired affective states.

Figure 1 shows the two components to the hybrid system that will help integrate the affective modelling foundations with the computational model. There are two components that are needed for full functionality of Tidal-MerzA.

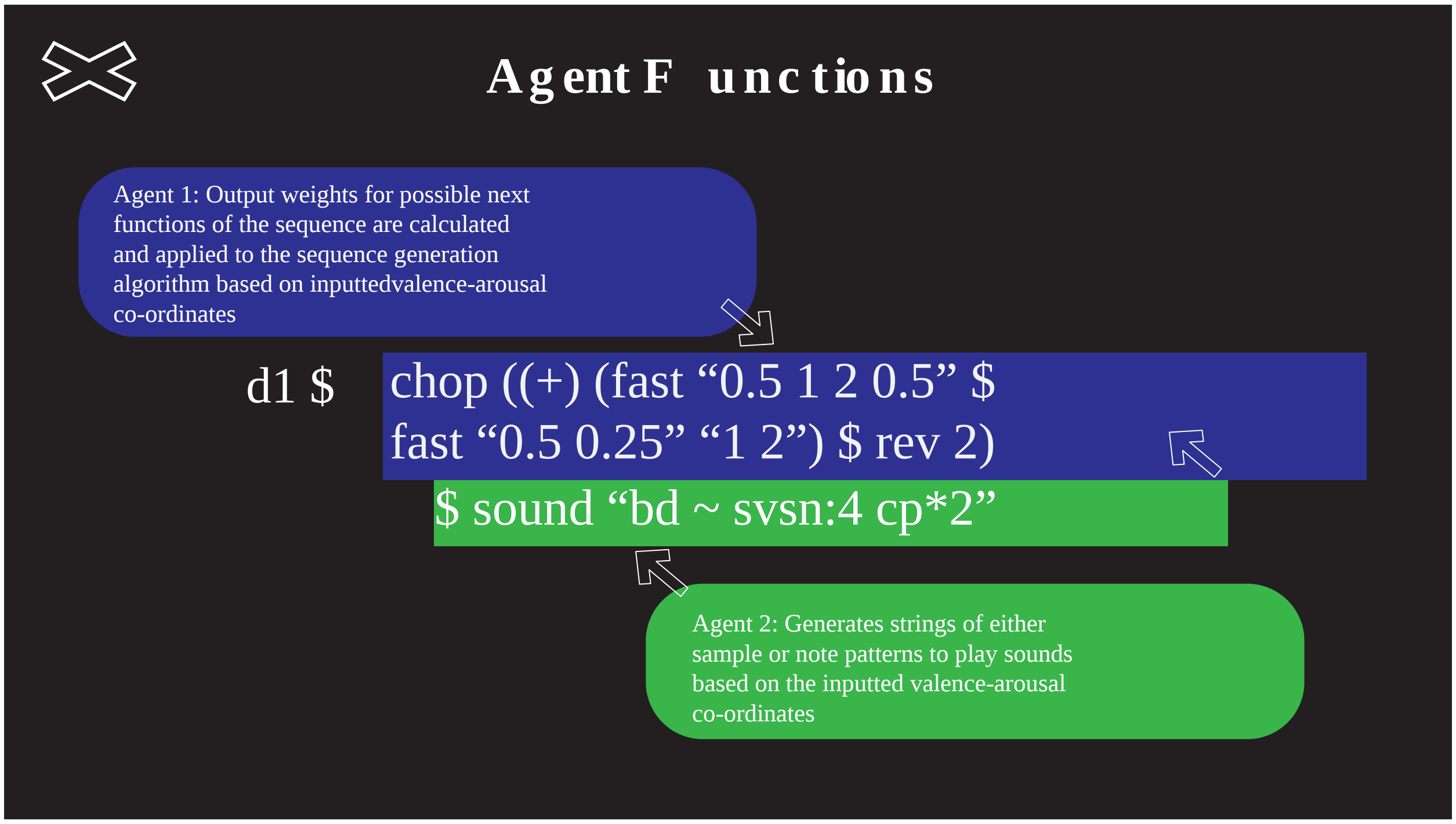

Figure 1 - Outline of how the two agents of the system work in conjunction to produce sequences of code in TidalCycles

Firstly, output weights for the Tidal-Fuzz agent are learnt based on valence-arousal co-ordinates by the design of the first agent. In Tidal-Fuzz, the selection of the next step in the sequence is determined by the weights calculated from an n-gram model based on user inputs. This gave a meaningful way to create the weights for the demonstration of how the agent should function. However, these weights are in a sense trivial and based on training based on a corpus of TidalCycles code, provided by the community (Wilson et al. 2021).

Secondly, another agent is used in the generation process. This second agent is in charge of creating the strings supplied to the `sound` function in TidalCycles. These strings define either melodic note sequences or rhythmic patterns to be played, based on either samples or synthesis in SuperCollider. These are the fundamental ways in which sound is made in TidalCycles. In Tidal-Fuzz, these were pre-written by the human and chosen at random. In Tidal-MerzA, these are generated from the ground-up by using another agent that learns based on the affective models.

The overall algorithmic structure of both this first agent can be seen in Figure 2 and the structure of the second agent in Figure 3.

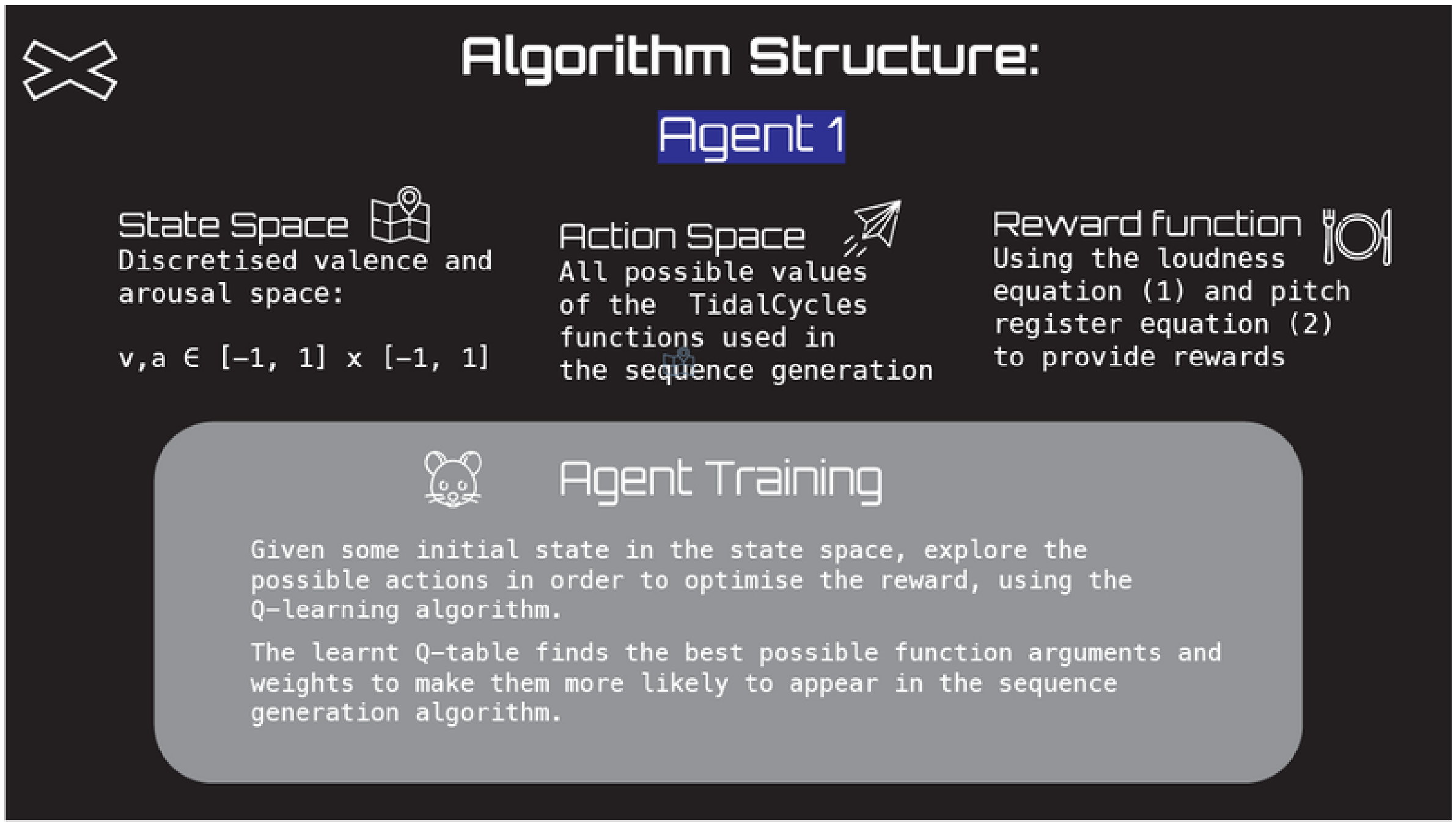

Figure 2 - Algorithm structure for the first agent in the MerzA system which uses reinforcement learning to learn structures for possible sequences of code

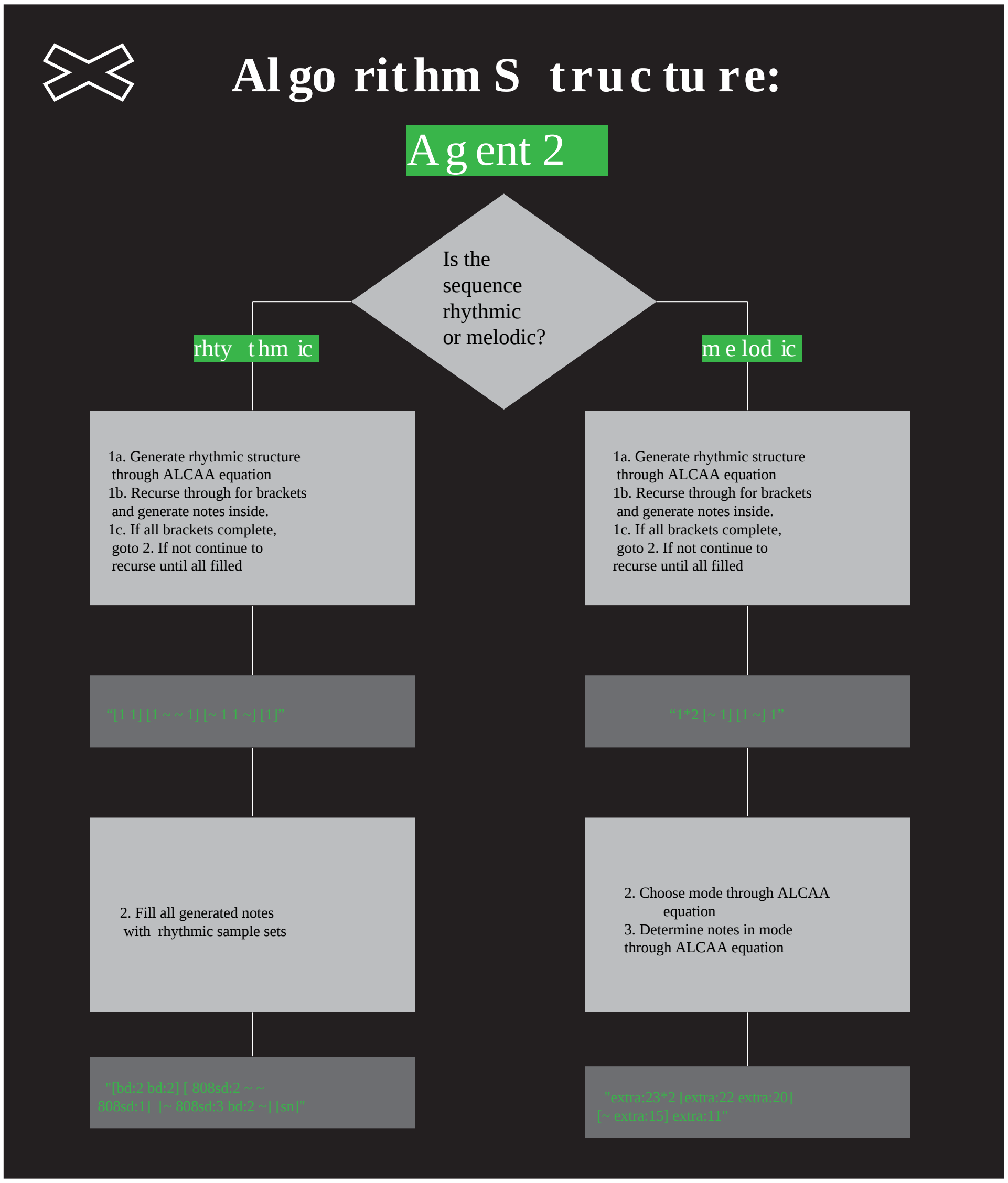

Figure 3 - Algorithm structure for the second agent in the MerzA system which uses the rule based system from ALCAA **(Wilson et al. 2024)** to create mini-notation strings

These musical structural parameters and how they were learnt through the agents can be seen in Figure 4. Notably, tempo was not included in either agent structure. This was due to the fact that tempo is usually controlled not through the Tidal patterns themselves, but through the function `setcps` that determines the cycles-per-second, and by extension, tempo. The next sections will detail how this hybrid system combined the modelling equations from ALCAA and computational model from Tidal-Fuzz as a way to leverage both into the generation of musical pattern in TidalCycles with specific affective qualities.

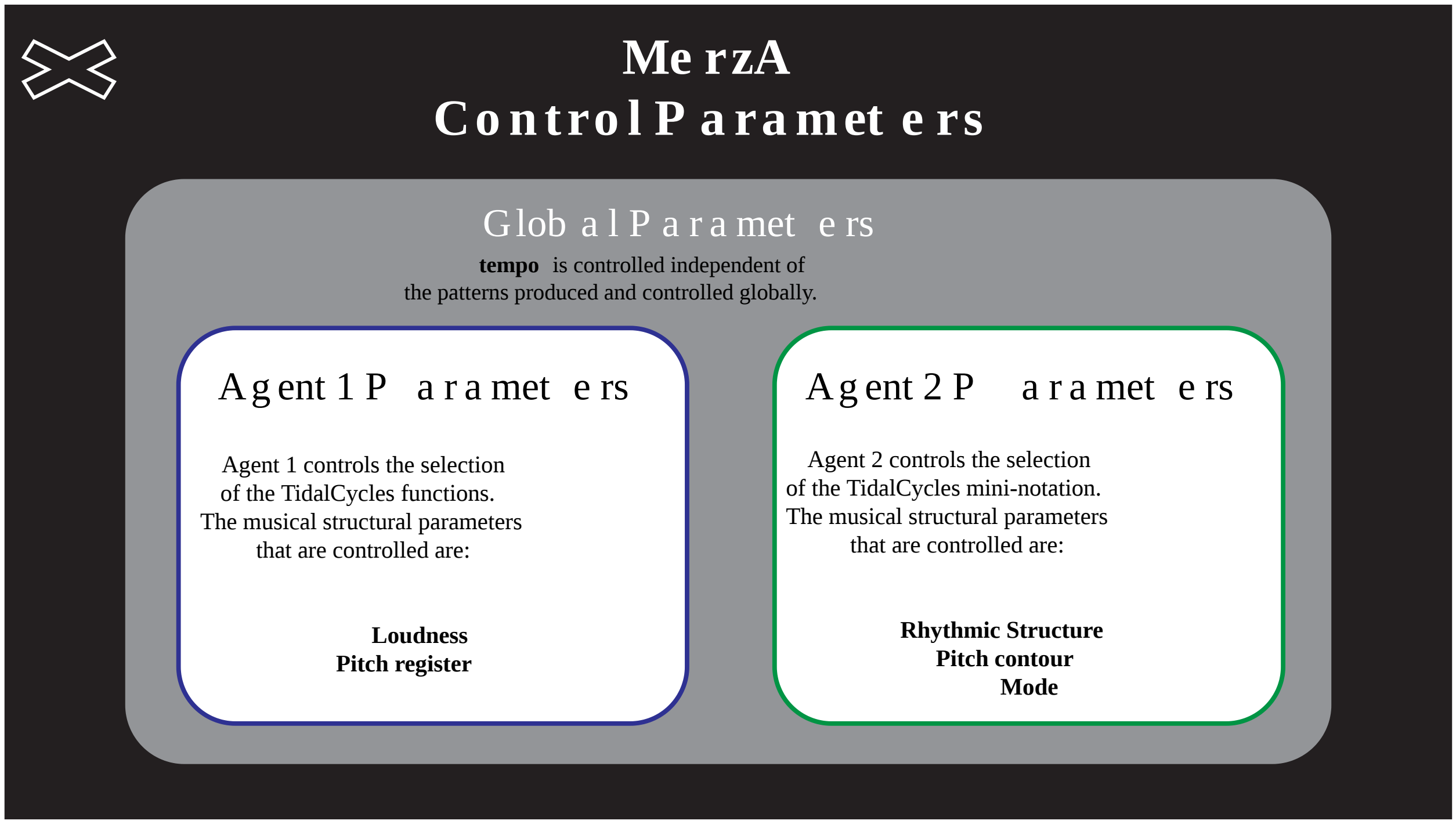

Figure 4 - An overview of the parameters of the affective model controlled by each agent.

## Design of Agent 1: Learning Weightings

The RL agent is designed using the Q-learning algorithm, a model-free, value-based reinforcement learning technique (Kaebling et al., 1996). The agent's objective is to learn a Q-function, which estimates the expected cumulative reward for taking a particular action in a given state. The Q-learning algorithm iteratively updates these Q-values based on observed rewards and the Bellman equation (Barto 1997).

### State Space

To represent the state, the agent utilises the valence-arousal coordinates that describe the desired affective qualities, i.e. $v, a \in [-1, 1] \times [-1, 1]$. The state space is discretised to represent the different possible states on this interval. The valence and arousal are both divided into ten segments, which gives a possible $10 \times 10$ state space, i.e. 100 possible states. This allowed to quantify the states with enough granularity, however limitations of this approach include potential loss nuance due to the discretisation process. A 2-dimensional-array where each row represents a unique combination of valence and arousal values is created to store all these states.

### Action Space

The action space for this RL agent is defined as the selection of sequences of code based on the valence-arousal coordinates. The agent discretised the action space to represent a set of predefined code sequences. This allows for flexible and fine-grained control over the generated musical code.

The design of the action space for MerzA encompassed the two main aspects: weightings for the `gain` function influenced by the loudness equation, and inputs for the `note` function based on the pitch register equation.

To define these aspects more clearly:

- Loudness: The loudness is defined by the following equation, based on the literature (Scherer 1977) (Ilie and Thompson 2006):
  - $l(a,v) = \text{unif}\{l_{min}, l_{min} + l_{range}\}$
    where
    $$\begin{cases} l_{min} & = -18 \\ l_{range} & = l_0 * a + l_1 * v \end{cases} \tag{1}$$
    Since the loudness is determined by a uniform distribution between $l_{\text{min}}$ and $l_{\text{min}} + l_{\text{range}}$ each action in this context can be a specific loudness level within this range. This range is discretised into a set of possible loudness levels of 25-equal intervals.
- Pitch register: The pitch register is determined by the given pitch equation based on the literature (Gabrielsson 2010):
  - $$p_r(a,v) = \begin{cases} \text{round}(v * 12), & \text{if } v > 0. \\ \begin{cases} \text{round}(\frac{(a-v)}{2} * 12)), & \text{if } a > 0. \\ \text{round}(\frac{(a+v)}{2} * 12)), & \text{if } a < 0. \end{cases}, & \text{if } v <= 0. \end{cases} \tag{2}$$
    $p_r(a,v)$ produces a value between 0 and 24. This range is also discretised into a set of 25 distinct values.

Since the agent needs to decide both the loudness and pitch register simultaneously, the combined action space would be the Cartesian product of the two individual action spaces. As they have been discretised each into 25 possible loudness levels and 25 distinct pitch levels, the combined action space has 625 possible actions.

With the state space of size $n$ and action space of size $m$ defined, the q-table then was initialised of size $n \times m$. In this case, the state space size $n = 100$ and the action space size $m = 625$ meant that the q-table is a matrix of size $100 \times 625$.

## Reward Functions

The reward function serves as a crucial guide for the agent, indicating the desirability of its actions and influencing its learning process. In the context of MerzA, defining appropriate reward functions was essential for shaping the system's ability to generate music that aligns with the specified affective dynamics.

The reward function was designed to encourage the desired behaviour, which in this case involved generating weightings for the values of the `gain` and `note` function arguments based on the given valence-arousal coordinates. Firstly, the agent received a reward based on how closely some chosen loudness matches the target loudness defined by the loudness equation. This encourages the agent to learn the appropriate loudness level for given valence-arousal coordinates. Similarly, the agent should be rewarded based on how appropriately it selects the pitch register according to the pitch register equation for the given affective state.

To construct such a reward function, firstly, the target loudness based on the loudness equation for given valence-arousal coordinates was calculated. The closer the agent's chosen loudness is to the target, the higher the reward. The reward function employed a negative absolute difference approach (Sutton 2018) to encourage precision in matching the target loudness and pitch register. Secondly, the target pitch register using the pitch register equation was determined and the agent was rewarded for selecting a pitch register close to the target. Again, this was done using a negative absolute difference.

One of the challenges encountered was ensuring that the agent did not exploit the reward system in ways that detracted from the overall musical quality. To address this, the reward structure was designed to consider not only the immediate outcome of an action but also its consistency with preceding and succeeding actions. The reward function calculates target values for loudness and pitch based on the current state, and then assesses the chosen action by computing the negative absolute difference between the chosen and target values for both loudness and pitch (e.g., `loudness_reward = -abs(chosen_loudness - target_loudness)`). This design ensures that the agent's decisions contribute positively to the overall flow and structure of the musical piece by discouraging large deviations from the targets. Other reward functions, such as

## Learning Algorithm and Exploration

The RL agent employed a Q-learning algorithm as the means with which it learnt and updated its decision-making policies based on the received rewards. Q-learning was chosen as it is a useful learning algorithm for problems with discrete action spaces (Barto 1997). Q-learning is an off-policy learner that aims to learn the value of the optimal policy, thus allowing the agent to evaluate the potential of actions without explicitly following them (Li 2023). This quality makes it particularly suited for environments where exhaustive exploration of the action space is impractical.

In the implementation in Agent 1 of Tidal-MerzA, the learning process revolved around updating the Q-learning table. Each entry in the Q-table represents an estimate of the expected cumulative future rewards for taking a given action in a given state, known as the Q-value.

The learning process unfolded over many episodes, where each agent represents a sequence of decisions. At every step within an episode, the agent observed its current state, selected an action based on either *exploration* or *exploitation*, and received a reward as the result. The reward reflected how well the chosen action contributed to achieving the desired affective outcomes in the music. The agent then updated the corresponding Q-value in the Q-table based on this reward, following the Q-learning update rule.

Exploration is crucial in the early stages of learning to ensure a diverse experience and to prevent the agent from converging prematurely to sub-optimal policies. On the other hand, exploitation involves choosing actions based on the current best knowledge, i.e., selecting actions with the highest Q-values in the Q-table. This approach enabled the agent to build upon and refine the successful strategies it had already discovered, gradually improving its ability to weight the function parameters to align with the specified affective dynamics. In the development of this RL agent, the exploration-exploitation balance was managed by an $\epsilon$-greedy policy, where $\epsilon$ is a parameter that determines the likelihood of taking a random action. Over time, $\epsilon$ was often decayed, gradually shifting the agent's behaviour from exploration-dominated to exploitation-dominated. This transition is key to the agent's ability to learn from its experiences and converge towards an optimal policy (Tokic and Palm 2011).

Through this iterative process, the RL agent progressively enhanced its decision making policies. This in turn led to continual improvement in its ability to select the values for the parameters that align with the ALCAA model. The Q-learning approach contributed to the agent's ability to learn and improve over time.

## Training the Agent

The training process of the RL agent is tailored towards its task of learning optimal strategies for weighting the functions based on valence-arousal co-ordinates. After defining the key components of our reinforcement learning model, the agent is set up so that it can actively interact with this environment. Specifically, the agent employs a function to determine the subsequent state it will occupy, which is dictated by its current state and the action it takes, and it also calculates the reward it receives for taking this action. Following each action executed by the agent, it updates its Q-table—a data structure used to estimate the expected rewards for each possible state-action pair. The update rule incorporated the received reward and discounted estimate of future rewards.

The agent then undergoes training over numerous episodes each comprising a series of steps until a termination condition is met. For this agent, it was found that 12000 episodes sufficed to produce results that were invariant to small changes in the input or environment. After this, there was a plateau in the rewards gained by the agent —which demonstrated that more learning would not produce more significant results. This level of training allowed the agent to achieve a stable and consistent performance, indicating a convergence of the learning process. During these episodes, the agent interacted with its environment, making decisions, receiving feedback in the form of rewards or penalties, and incrementally improving its policy based on this feedback.

Once the training was completed by the agent, it can then determine the optimal gain and note settings for any given valence-arousal pair.This is achieved by querying the trained Q-table with the valence-arousal state and retrieving the action that maximises the expected reward. By the end of the training process, the agent is capable of adapting its outputs to the range of valence-arousal co-ordinates for $v, a \in [-1, 1] \times [-1, 1]$.

Multiple training sessions were conducted, each time gathering data, analysing the agent's performance, and making any necessary adjustments to the hyperparameters. This iteration process was key in making sure that the number of episodes, total number of steps in each episode, and hyperparameters were configured correctly to adjust the agent's ability to create music that aligned with the specified affective dynamics. Specifically, the hyper-parameters of the learning rate, discount factor, and $\epsilon$-greedy parameter were fine-tuned to optimise the learning process. Through this process of continual evaluation and iteration, Agent 1 evolved into a more adept system.

# Design of Agent 2: Mini-notation Strings

As outlined, there were two aspects that were needed to create executable patterns of TidalCycles code. As outlined in Figure 1, a second agent was used to generate the mini-notation strings for the code sequences. This task involved the synthesis of sequences of tokens to form coherent and contextually relevant strings. In (Wilson et al. 2021), one challenge was noted in the generation of these strings. Specifically, the mini-notation is a terse way to represent events within a pattern in Tidal syntax, and these additional complexities of notation were omitted in this early version, where mini-notation strings are treated as single tokens.

In response to this challenge and to generate notation events, a novel agent is proposed that generates mini-notation strings through dynamic adjustment to the importance of individual tokens, using the affective model outlined in (Wilson et al. 2024). This section explores the architecture and experimental outcomes of this second agent in combination with the first.

In this section, as outlined by Figure 3, the mode, pitch contour and rhythmic structure is determined through the mini-notation, rather than learning function transformers, as seen in the design of agent 1. Internally, the mini-notation is actually parsed and understood as a shortcut for a function that you could otherwise write using longer function compositions. The mini-notation is used for this agent as this allows the generation of new strings as this is an easier abstraction to work with than the functions themselves. The previous agent outlined uses function the reinforcement agent to learn weightings as these functions are not expressible in the mini-notation.

## Rhythmic Roughness

Generation of rhythmic and melodic sequences are treated separately, with generation of rhythmical structure used to give structure to the melodic patterns, in a similar manner to other affective algorithms (Morreale and De Angeli 2016) (Ehrlich et al. 2019) (Agres 2023). The generation of these sequences will be treated in a different manner, which are now outlined.

Overall, the agent acts in the following manner for rhythmic sequences. Firstly, the possible rhythmic structure tokens, T, that can be generated is defined as:

$$\mathrm{T} = [`\sim', `1', `1{*}2', `1@2', `[]']$$

These form a subset of the whole mini-notation.

This subset was chosen as it allows the representation of all the concepts needed rhythmic roughness described in (Gundlach 1935) . The token `1` represents a sound occurring, `~` represents a rest, `1*2` repeats the note in the same segment (i.e. creates quavers from a crochet, semiquavers from quavers, depending on the number of events in the sequence). The `1@2` token elongates a pattern (i.e. the inverse, creates a crochet where notes are divided into quavers etc). And finally the `[]` creates a pattern grouping. Similar to `1*2`, this shortens the length of this element. However it does not just repeat the previous pattern but allows sub-groupings of all the previous token types.

The algorithm for generating mini-notation strings for the rhythmic patterns is then outlined as follows:

1. Construct a sequence of fixed length from the possible states, T, based on the valence and arousal input parameters
2. Once the first sequence is completed, check the string using regular expression to see if any bracket tokens are chosen
3. If brackets exist, move inside and return to step 1. If there are no brackets in the sequence, go-to step 4
4. From a predetermined set of drum samples, randomly select a sample to replace all 1s in the sequence.

The equation for generating rhythmic patterns, from (Wilson et al. 2024) , is outlined as follows:

$$R(a, n) = \frac{(1-a)*(n+1)}{2}$$

where $R(a, n) = Pr(\text{note removed})$.

The transformation into TidalCycles mini-notation from this equation is thus as follows. Roughness is a parameter used to determine the variation in note lengths over a measure of music: if all notes are of equal duration, roughness is low, and if notes are of varying length, roughness is high (Gundlach 1935). This formula is used to determine the roughness parameter, based on arousal input, by calculated probabilities to select a token that will either increase roughness or decrease it.

From this, note patterns of set length can be generated as mini-notation pattern strings, based on the input arousal parameter, $a \in [-1, 1]$.

For example, where $a = 0.65$, and using a sample called `kick` the mini-notation pattern that is generated is:

```
”kick kick*2 ~ [~ kick] ~ kick ~ kick*2”
```

or where $a = -0.25$, and different samples, the pattern that is generated is:

`“bd ~ 808oh bd sd sd sd [~ 808oh]”`

## Modality

The next part of the creation of mini-notation strings is to select the mode based on the valence parameter. This mode provides a key profile for which to generate the pattern from.

The equation for determining the mode (Wilson et al. 2024) , or some ordering of the modes, outlined in (Schmuckler 1989):

$$\begin{aligned} M = [\ & \text{“lydian”}, \text{“ionian”}, \ldots \\ & \ldots \text{“mixolydian”}, \text{“dorian”}, \text{“aeolian”}, \ldots \\ & \ldots \text{“phrygian”}, \text{“locrian”}\ ] \end{aligned}$$

then, based on the valence parameter, the index of the matrix $M$ is chosen following the equation, according to (Ehrlich et al. 2019):

$$m(v) = M[\text{round}(3 - 3.5v)]$$

This provides the modality for the agent. Again, this rule-based system can be applied to mini-notation strings for this agent.

To do this, requires the samples of each folder to be ordered chromatically, where `"sample:0"` represents the root note, `"sample:1"` represents one semitone above and so on. Then, following this ordering of the samples, a dictionary can be constructed for each of the modes as following:

```
modes = {
 “lydian”: [0, 2, 4, 6, 7, 9, 11, 12],
 “ionian”: [0, 2, 4, 5, 7, 9, 11, 12],
 “mixolydian”: [0, 2, 4, 5, 7, 9, 10, 12],
 “dorian”: [0, 2, 3, 5, 7, 9, 10, 12],
 “aeolian”: [0, 2, 3, 5, 7, 8, 10, 12],
 “phrygian”: [0, 1, 3, 5, 7, 8, 10, 12],
“locrain”: [0, 1, 3, 5, 6, 8, 10, 12]
}
```

As before, the code initialises an argument parser to accept valence and arousal parameters from the command line, with both parameters defaulting to 0.5 if none are provided. Then, a rhythmic patterns of tokens are generated in the manner formerly described. The valence parameter influences the choice of musical mode, with the script supporting various modes like Lydian, Ionian, and others, each defined by specific note intervals. The arousal parameter, on the other hand, influences the rhythmic complexity of the generated sequence. It determines the probability of rests occurring in the music and affects the likelihood of different rhythmic patterns, defined as tokens like `1`, `1*2`, and `1@2`, with each token representing a rhythmic element or a rest. Overall, the algorithm selects a mode based on the valence, calculates token probabilities based on arousal, and generates a musical sequence. Using regular expressions, it can then apply the degrees of the scale to the sample library selected. However the order with which the samples are chosen is based on the final parameter, pitch contour.

## Pitch contour

The next part of the creation of mini-notation strings is to select the degrees of the mode for each note, based on the valence parameter. The mode chosen, as formerly outlined, provides a key profile for which to generate the pattern from.

It was outlined in (Gabrielsson and Linström 2010) that ascending melodies are associated with positive emotional states (i.e. high valence affect), whereas descending melodies are associated with negative emotional states (i.e., low valence affect). This finding was modelled through the equation:

$$p_c(v, i, j) = \begin{cases} \frac{v+1}{2} \cdot w(K[j] - K[i]), \text{if } v > 0 \\ \frac{1-v}{2} \cdot w(K[j] - K[i]), \text{if } v \leq 0 \end{cases}$$

The final stage for producing the mini-notation strings for the melody involved selecting the degrees of the chosen mode using this equation and then selecting the corresponding sample for the chosen mode, i.e. given a probability that the next note will be higher in the sequence, with this probability being modelled using $p_c(v, i, j)$. This meant that for a higher valence, there was a greater chance of ascending to a higher note in the mode, and a greater chance of descending for negative valence.

As this is applied to the generated rhythmical sequence, fixed length note patterns can be generated as mini-notation pattern strings, based on the input arousal parameter, $a \in [-1, 1]$. As an example, for the valence-arousal co-ordinates $v = 0.8$ and $a = 0.65$, and with a sawtooth sample `"saw"` selected, the generated melody would be:

```
"~ saw:4 saw:6*2 saw:7 ~ saw:9 saw:11 saw:12*2"
```

Likewise, for $v = -0.25, a = -0.8$:

```
"~ saw:10*2 ~ saw:8 ~ ~ saw:7 ~"
```

# Performing with Tidal-MerzA

The hybrid system described here produces both the weightings and mini-notation strings for the live coding agent. These are then used in conjunction with an existing auto-complete Atom plugin that was outlined in (Wilson et al. 2021). However, due to the sunsetting of the Atom Text-Editor since this work was completed, this was remade as a plug-in for the Pulsar open-source text editor.  This custom plugin is combined with a Haskell listener module that requests a pattern when a `$` command from the Atom Editor is executed. This allows the live coder to receive a completed pattern to accept or reject. The format of this for MerzA was similar, entailing the use of the Atom plugin to create new patterns  on receipt of a `$` symbol in the editor.

However, a slight difference was that first the agent needed to be given a valence and arousal parameter. These were currently sent from the command line using argument parsing in python. Once these had been received, the training completes and MerzA outputs a text file with the learnt function weights and mini-notation strings was produced.

This file was formatted in the same way the n-gram models previously were, as arrays of tuples with normalised weights, in the manner outlined in (Wilson et al. 2021). The learnt weights and mini-notation strings were merged with the file from the previous agent structure. A listener function was created for this system, so that once the co-ordinates were received and the training had been completed, patterns were automatically suggested in the text editor.

The process of training and using the agent can be seen here:

https://www.youtube.com/watch?v=QBl3c7wWWPU

# Evaluating Outcomes

Firstly, the learning efficiency of the RL agent—Agent 1—is discussed. Through the use of a tracking mechanism in the training loop, this allowed the monitoring of the rate of improvement in rewards over episodes. After training, a plot keeping track of the moving average for the reward function was produced, seen in Figure 5. This approach allowed quantitative evaluation of the learning efficiency of the agent.

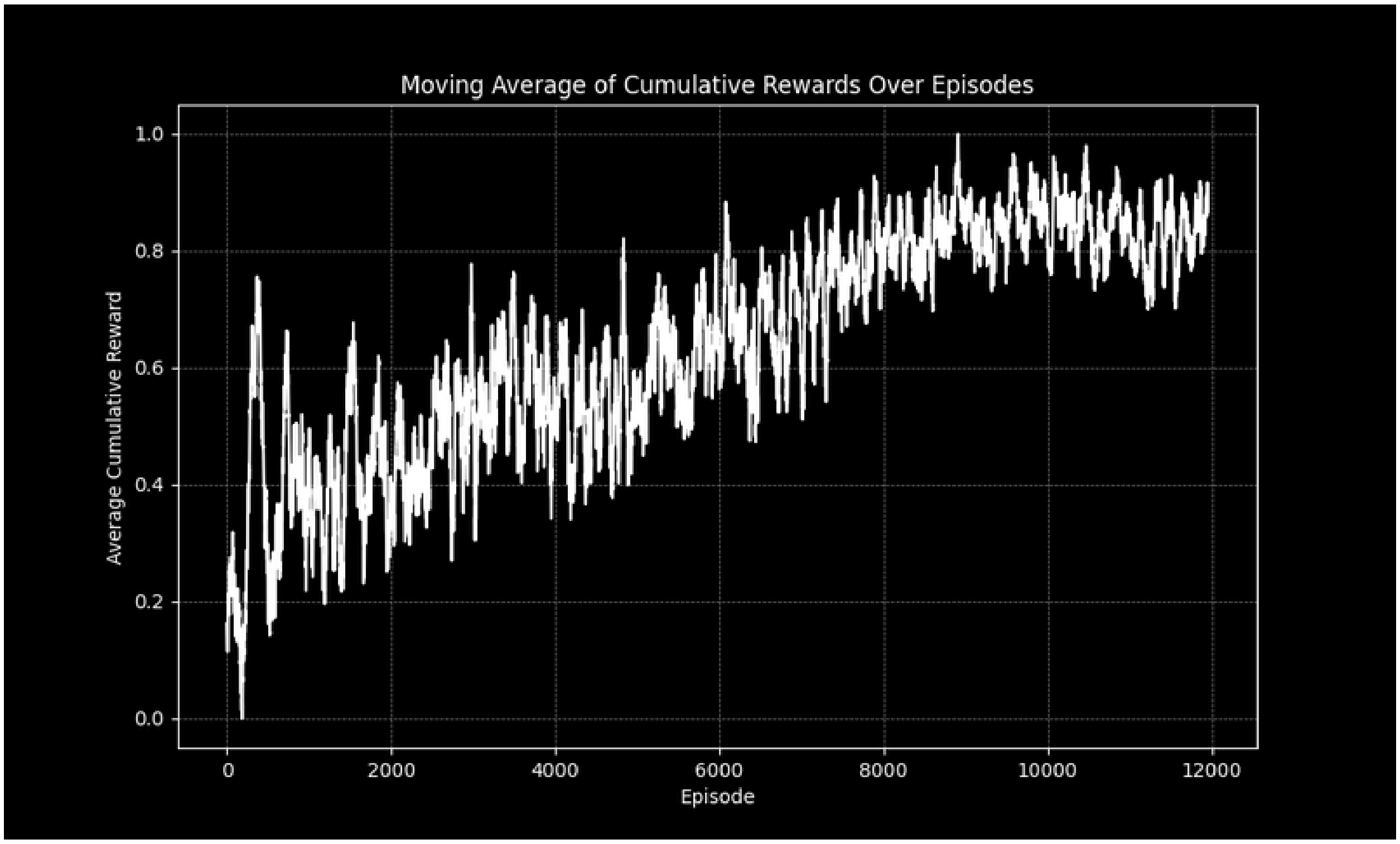

Figure 5 - Overall cumulative reward function for Agent 1 in the Hybrid System

In Figure 5, an upward trend in the plot of cumulative rewards over episodes can be observed. This indicates that the agent is learning and improving its performance over time, achieving higher rewards in later episodes. The large fluctuations in the initial episodes indicate that the agent is exploring different strategies.

The overall aim of generating music that conveys specific affective qualities through modelling musical structure has already been evaluated through quantitative testing (Wilson et al. 2024). As this agent used the same model in the generation process, this is not evaluated again.

Subsequent investigations will focus on evaluating the agent's usefulness in a performance context, where new evaluation methods are employed that will aim to evaluate the hybrid system itself as a means with which to further expand the live coder's creative strategies, but are outside the scope of this research.

## Advantages and Limitations of MerzA

Overall, MerzA successfully integrated the affective response model tested in (Wilson et al. 2024) with the computational model evaluated in (Wilson et al. 2021).  Through the outlined algorithms, it has been shown how co-ordinates on the valence-arousal affective space can be translated to patterns of TidalCycles code.

Using reinforcement learning in this context, combined with the mini-notation string generation algorithm, offers several advantages, especially in the context of adaptability, complexity, and creativity.

Firstly, it has the ability to learn from experience. This is useful as it provides some long term adaptability that the previous models did not provide. Further, the agent can continue to improve its performance over time through continued learning, allowing for refinement and optimisation of musical outputs based on accumulated experiences. Secondly, the outlined RL framework allows for the integration and balancing of the multiple objectives required by this task. Whilst at the moment it only allows the modelling of the specific equations outlined first in the ALCAA system, future iterations of this project could expand this to incorporate a wider scope.

There are limitations that currently exist with this system still. Whilst incorporating the `gain` and `note` functions, alongside representing melody and rhythmical pattern ideas in the mini-notation, was a step in the direction of generating autonomous musical pattern, there are still a wide range of pattern transformation functions that exist in TidalCycles that have not yet been incorporated into this model.

TidalCycles is particularly suited for the task of pattern manipulation, by its functional nature and its compact syntax representation. However, the relationships between pattern transformations and affective qualities are not currently well known. There exists literature on this for the visual arts (Takahashi et al. 2012) (Bertamini 2013) (Pecchinenda 2014) but for musical pattern this work is sparse. This gap indicates a need for further exploration to fully harness TidalCycles' capabilities in aligning musical patterns with desired affective states.

# Conclusion

In conclusion, this research has demonstrated the potential of using reinforcement learning to facilitate the learning of function weightings in computational music generation. The application of this technique has proven to be a step in advancing the field of affective algorithmic composition, showcasing a novel approach to integrating artificial intelligence algorithms into creative practices. Furthermore, through this hybrid system, mini-notation strings which are a key feature of Tidal's syntax, has enabled a concise expression of musical ideas. This development is significant as it allows the coding agent to select and manipulate musical concepts with greater precision and relevance.

The outcomes of this work highlight not only the technical feasibility but also the artistic potential of employing advanced computational methods in music generation. By bridging the gap between computational models and creative expression, this research paves the way for more sophisticated and nuanced musical creations, driven by intelligent computational systems.

Overall, the finding presented in this paper contribute insights to the field of computational music and open up new avenues for exploration in the intersection of artificial intelligence, creativity, and artistic expression. This chapter outlines how to fuse the work from the previous system developments. The next stage of development is to critically evaluate its outcomes and what this can mean for human-machine creative relationships.

## Acknowledgments

This work was supported by EPSRC and AHRC under the EP/L01632X/1 (Centre for Doctoral Training in Media and Arts Technology) grant. G. Wiggins received funding from the Flemish Government under the “Onderzoeksprogramma Artificiële Intelligentie (AI) Vlaanderen”.

## Ethics Statement

Overall, this work aims to adhere to ethical best practices for conducting research in the field of artificial intelligence and musical creativity.

In particular, the main ethical considerations for this work are:

(1) **- Issues around training data based on users musical creative outputs**. For the model to construct sequences of code, it was trained on a corpus of data. This data was provided by TidalCycles users through an open call to the community (Wilson et al. 2021). Participants were informed of how the data was to be used and were explicitly given the option to consent to providing data to be incorporated into the corpus.

(2) - **The potential for algorithmic bias and the representation of diverse musical genres and styles.** Given the reliance on user-provided data, there is a significant concern that the resulting AI model might inadvertently favour certain musical genres, styles, or techniques over others, depending on the diversity of the dataset. To mitigate this, efforts were made to ensure a broad and inclusive collection of users provided the corpus on which the agent was trained, however further analysis or collection of data would be useful.

(3) **- Intellectual property rights and the ownership of AI-generated music.** This work follows guidelines to respect intellectual property rights, including crediting original creators where possible.